\documentclass[journal,twoside]{IEEEtran}

\usepackage{amssymb,bm,mathtools}
\usepackage{amsmath}
\allowdisplaybreaks
\usepackage{epsfig,subfigure,colortbl,psfrag,subfloat}
\usepackage{enumerate,array,multirow}
\usepackage{cite,url}
\usepackage{algorithmic}
\usepackage{algorithm,setspace}
\usepackage{framed}
\usepackage{stfloats}% long eq
\usepackage{textcomp}
\def \diag{\operatornamewithlimits{diag}}

\def \tr{\operatornamewithlimits{tr}}
\def \min{\operatornamewithlimits{minimize}}

\def \ln{\operatorname{ln}}
\def \log{\operatorname{log}}
\def \exp{\operatorname{exp}}

\def \E{\operatorname{E}}

\def \rank{\operatornamewithlimits{rank}}

\def \dBm{\operatorname{dBm}}

\def \MHz{\operatorname{MHz}}
\def \Hz{\operatorname{Hz}}

\def \rank{\operatornamewithlimits{rank}}

\def \vec{{\rm{vec}}}
\def \beq {\begin{equation} }
\def \eeq {\end{equation} }
\def \beqn{\begin{eqnarray} }
\def \eeqn{\end{eqnarray} }
\def \bmat{\begin{bmatrix}}
\def \emat{\end{bmatrix}}
\def \bmats{\left[\begin{smallmatrix}}
\def \emats{\end{smallmatrix}\right]}
\def \beqi{\begin{IEEEeqnarray}{rcl}\IEEEyesnumber}
\def \eeqi{\end{IEEEeqnarray}}

\newtheorem{remark}{Remark}

\newtheorem{property}{Property}

\psfull

\begin{document}

\title{Principal Component Analysis (PCA)-based Massive-MIMO Channel Feedback}

\author{Jingon~Joung, Ernest~Kurniawan, and~Sumei~Sun% <-this % stops a space
\thanks{Part of this work has been published in Proceedings of the IEEE Globecom 2014, Austin, TX, USA, December 2014 \cite{JoSu14GC}.}
\thanks{The authors are with the Institute for Infocomm Research (I$^2$R), A$^\star$STAR, Singapore 138632 (e-mail: \{jgjoung, ekurniawan, sunsm\}@i2r.a-star.edu.sg)}%; $^*$J. Joung is the corresponding author.}
}
\maketitle

\begin{abstract}% 150-250 words: 240 words now
Channel-state-information (CSI) feedback methods are considered, especially for massive or very large-scale multiple-input multiple-output (MIMO) systems. To extract essential information from the CSI without redundancy that arises from the highly correlated antennas, a receiver transforms (sparsifies) a correlated CSI vector to an uncorrelated sparse CSI vector by using a Karhunen-Lo\`{e}ve transform (KLT) matrix that consists of the eigen vectors of covariance matrix (CM) of CSI vector and feeds back the essential components of the sparse CSI, i.e., a principal component analysis method. A transmitter then recovers the original CSI through the inverse transformation of the feedback vector. Herein, to obtain the CM at transceiver, we derive analytically the CM of spatially correlated Rayleigh fading channels based on its statistics including transmit antennas' and receive antennas' {\it correlation matrices}, channel {\it variance}, and channel {\it delay profile}. With the knowledge of the channel statistics, the transceiver can readily obtain the CM and KLT matrix. Compression feedback error and bit-error-rate performance of the proposed method are analyzed. Numerical results verify that the proposed method is promising, which reduces significantly the feedback overhead of the massive-MIMO systems with marginal performance degradation from full-CSI feedback (e.g., feedback amount reduction by $80\%$, i.e., $\frac{1}{5}$ of original CSI, with spectral efficiency reduction by only $2\%$). Furthermore, we show numerically that, for a given limited feedback amount, we can find the optimal number of transmit antennas to achieve the largest spectral efficiency, which is a new design framework.
\end{abstract}


\section{Introduction}

%\textcolor[rgb]{1.00,0.00,0.00}{Tx-Rx $\rightarrow$ BS-UE/MS}
In communications, channel state information (CSI) can be used at a transmitter (Tx) to improve a quality-of-service (QoS). The CSI at the Tx (CSIT) enables a preprocessing at the Tx to overcome a poor channel condition incurring severe performance degradation. Especially, preprocessing techniques using multiple transmit antennas, such as a single user multiple-input multiple-output (MIMO) beamforming and a multiuser (MU) MIMO precoding (see \cite{ZhGi02,SpSwHa04,ChMu04,JoKiLiJaShChChLe06,SaTaSa07,LiChGeSaShZh12} and references therein), are typical, promising methods for high-QoS communications systems.

The CSIT can be typically realized either by uplink (from receiver (Rx) to  Tx) channel estimation at a Tx in time-division duplex (TDD) systems, e.g., implicit feedback in IEEE 802.11n \cite{11n}, or by CSI feedback from an Rx to a Tx in frequency-division duplex (FDD) systems, e.g., explicit feedback in 802.11ac \cite{11ac}. Phase calibration improves the reciprocity between uplink and downlink (from Tx to Rx) TDD channels, so that the Tx can obtain the downlink CSI from the uplink CSI estimates \cite{GuKa13}. On the other hand, channel feedback of FDD systems has a fundamental issue on the channel feedback overhead against the limited uplink channel capacity. Especially, when the number of antennas is very large, possibly a few tens or hundreds of antennas \cite{BjHoKoDe13,RuPeLaLaMaEdTu13}, i.e., a massive or very large-scale MIMO, the channel feedback overhead issue becomes more severe and it renders the closed-loop feedback approach impractical. Hence, network throughput improvement cannot be guaranteed due to the uplink overhead even if the downlink throughput is improved \cite{JaJoShJe11}.

\IEEEpubidadjcol

For a conventional MIMO system with usually less than $10$ antennas, reduced feedback information has been rigorously studied, such as a codebook (see \cite{YiVoThLoGh14} and references therein), channel distribution \cite{GhRaZe10}, partial CSI \cite{MaEr10}, and implicit CSI (e.g., rank, precoding matrix, and channel quality indicators) in \cite{LiChGeSaShZh12}. However, it is difficult to directly apply the schemes to a massive MIMO system. For example, the codebook in general requires high computational complexity, even for the case of eight transmit antennas \cite{R1-090618}. The high complexity design issue also persist in a random vector quantization (RVQ) approach for the codebook generation. Hence, a reduced size of codebook can be considered for the systems.  However, due to the very large dimension of the massive MIMO channels, this approach will degrade the communication performance severely \cite{WaCoDeSl12}, and the optimal finite-size codebook design for the very large dimension arises as a new issue. Another issue related to method is the scalability for an MU massive MIMO scenario. The massive MIMO may not embrace the optimality loss of direct design of MU-MIMO codebook from single-user MIMO codebook, e.g., the 3rd generation partnership project long-term evolution Release-8 MU-MIMO codebook to reduce the channel feedback amount, we may consider distributing the many antennas and feeding back only the local CSI for strong channels instead of the global CSI including weak channels \cite{JoChSu14}. As a directly possible way for collocated massive MIMO systems, CSI (or equivalent antennas) grouping is proposed \cite{LeSh14,LeChSeLoSh15}. As demonstrated in \cite{LeChSeLoSh15}, correlation due to the many collocated antennas within the limited space of massive MIMO Tx and Rx imposes redundancy on CSI information and it makes possible that the amount of feedback can be significantly reduced by grouping the highly correlated CSIs, although the best grouping pattern and the total number of groups are yet to be analyzed further \cite{LeSh14}.

In this work, we consider {\it compressive channel feedback} using a sparse {\it principal component analysis} (PCA) technique, and propose a sparse channel feedback (SCF) method based on the PCA for massive MIMO systems. The PCA is a well established tool for various applications, such as genetics, chemistry, meteorology, image processing, machine learning, and data mining, to reduce high dimensional {data} to a smaller dimension by also exploiting the correlation in the data. Precisely, the PCA extracts $M$ principal components that are uncorrelated from $N$ correlated components ($N\gg M$) by using signal transformation \cite{Jo86Book,JoNeRiSe10}, i.e., a dimensionality reduction of a set with correlated components. For the transformation, the PCA employs optimal transformation using a Karhunen-Lo\`{e}ve transform (KLT) matrix that consists of eigen vectors of a CSI covariance matrix such that the original CSI can be compressed efficiently with no correlation\footnote{Note that PCA exploits the correlation information between CSIs to reduce the feedback amount, yet it achieves the compression by representing the CSI vector in terms of its dominant eigen spaces instead of grouping the CSI with similar values as in \cite{LeSh14}.} \cite{LiCh93,KuKuTi12,GwKuVl12,KuJoSu15}. The Tx can recover the original CSI by inverse-transforming the sparse CSI with the KLT matrix based on a compressive sampling/sensing (CS) theory \cite{CaRoTa06,MaQuMuRoWiZo09,KuKuTi12}. Notwithstanding the optimality of the KLT in terms of the compression, it is challenging to employ the KLT in the channel feedback due to the data/signal-dependent characteristic of the KLT matrix. Introducing alternative methods, such as an adaptive algorithm\footnote{Recently, in \cite{KuJoSu15}, a tracking algorithm for channel's principal component is proposed by tracking a perturbation term of CSI.} \cite{KuKuTi12,KuJoSu15} and an empirical method \cite{GwKuVl12}, we derive a closed form expression of the {\it covariance matrix} of spatially correlated Rayleigh fading channels, which consists of transmit and receive {\it correlation matrices}, channel {\it variance}, and channel {\it delay profile}. The tractable representation of the channel covariance matrix enables the implementation of PCA-based channel feedback and provides analytical tradeoff between data compression and communication performance, i.e., a normalized mean-squared error (NMSE).
%Consequently, the KLT matrix can be fixed and shared between the Tx and Rx in offline mode before communications, and the SCF can be readily implemented in the communications.
Numerical results verify that the proposed SCF method can improve a compression performance, i.e., increase a compression ratio sustaining communication performance. For example, feedback amount can be reduced by $68\%$ with no spectral efficiency loss and it can be reduced by $80\%$, i.e., $1/5$ amount of original CSI, with spectral efficiency reduction by only $2\%$. Furthermore, the SCF method can have lower implementation complexity and, in practice, could be more stable and robust as compared {to the} instantaneous channel feedback schemes. At last, we justify that the proposed SCF method using KLT is a promising channel feedback method for the massive MIMO systems by answering for one possible question ``Why do not we just reduce the number of transmit antennas and feed back the full CSI without compression?''

The rest of the paper is organized as follows. In Section \ref{Sec:CF-CS}, channel feedback and compressive sampling are briefly introduced. In Section \ref{Sec:SCF}, we propose an SCF method using PCA. Section \ref{Sec:Error-Analysis} provides error analysis on compressive feedback. Simulation results are shown in Section \ref{Sec:Sim}. Section \ref{Sec:Conclusion} concludes the paper. %The specific 

\underline{Notations}: $\|{\bm a}\|_p$ represents $p$-norm of vector ${\bm a}$; ${\bm A}^{-1}$, ${\bm A}^{\dag}$, ${\bm A}^T$ and ${\bm A}^H$ are the inverse, pseudoinverse, transpose, and Hermitian transpose of ${\bm A}$, respectively; ${\bm I}_a$ and ${\bm 0}_{a,b}$ are $a$-dimensional identity matrix and an $a$-by-$b$ zero matrix, respectively.

\section{CSI Feedback and Compressive Sampling}\label{Sec:CF-CS}
We first introduce the model of channel feedback, and briefly recapitulate the basic of compressive sampling/sensing (CS) to interpret the channel feedback method from CS perspective. The interpretation of channel feedback based on CS will help us to understand the sparse-domain channels and the sparse channel feedback (SCF) mechanism, and to capture the essential part of the proposed SCF in Section \ref{Sec:SCF}.

\subsection{CSI Feedback}
Let a channel vector be ${\bm h}\in\mathbb{C}^{N\times 1}$ that consists of spatial-and-frequency domain $N$ channel elements and is supposed to be fed back to a Tx. When we compress $N$ samples of ${\bm h}$ to $M$ samples, where $M\leq N$, a data (information) compression ratio is defined for given $N$ as
\beq\nonumber%\label{CSratio}
1\leq \gamma(M)\triangleq{N}/{M}<\infty.
\eeq

In this study, we consider a {\it compressive feedback error} that arises only from the compression not from estimation at an Rx and Tx. In other words, we assume that an Rx can estimate and feed back the compressed information of ${\bm h}$ perfectly, and a Tx can also obtain the compressed $M$ feedback samples without distortion. Hence, the discrepancy between the recovered channels $\widetilde{\bm h}$ at the Tx and the original channel ${\bm h}$ may exist only if $\gamma(M) > 1$, and it is quantitatively measured by a normalized mean-squared error (NMSE) $\delta(M)$, i.e., a compressive feedback error, defined as 
\beq\label{NMSE}
0\leq \delta(M) \triangleq {\E\|{\bm h}-\widetilde{\bm h}\|_2^2}\Big/{\E\|{\bm  h}\|_2^2}<\infty.
\eeq
We will derive $\delta(M)$ analytically in Section \ref{Sec:Error-Analysis}.

Larger $\gamma(M)$ and smaller $\delta(M)$ are desired for {\it efficient} (i.e., less feedback overhead sustaining performance) and {\it reliable} (i.e., less performance degradation) communications. Typically, there is a tradeoff between the efficiency and the reliability in communications (or tradeoff between interpretability and statistical fidelity in data acquisition \cite{JoNeRiSe10}). In other words, if $M$ decreases (or increases), both compression ratio $\gamma(M)$ and NMSE $\delta(M)$ increases (or decreases). However, the tradeoff may disappear depending on the channel characteristics. For example, we can increase $\gamma(M)$ sustaining the $\delta(M)$ if ${\bm h}=[h\cdots h]^T$. In  {that} case, a Tx can achieve a zero NMSE from feedback of only $h$ for any $\gamma$. Note that the basic assumption in the example is that the Tx knows that the channel is {\it static}. The lesson from the example is that the knowledge of the channel statistics, such as correlation information in the example, can be used to reduce CSI, which inspires us to consider an SCF method based on CS. %Before we jump into directly the proposing SCF, we briefly recapitulate a compressive sampling/sensing (CS), and interpret the channel feedback (channel feedback) method from CS perspective, which will help to capture the essential part of the proposed SCF.

\subsection{Compressive Sampling/Sensing (CS)}
The CS is a technique to recover $N$ original samples (i.e., ${\bm  h}$) from its compressed {$M<N$} observation samples (i.e.,  {${\bm y}\in\mathbb{C}^{M\times1}$}) \cite{CaRoTa06,MaQuMuRoWiZo09,KuKuTi12}. To extract the effective information from ${\bm h}$ and to construct an observation vector ${\bm y}$, a {\it measurement} matrix  {${\bf \Phi}\in\mathbb{C}^{M\times N}$} is used as
%\beq\nonumber
${\bm y}={\bf \Phi}{\bm h}.% = {\bf \Phi}{\bf \Psi}^{-1}{\bm s}
$ %\eeq
Now, the CS forms $\ell_1$ minimization problem as
\beq\label{CS}
\widetilde{\bm s} = \min_{{\bm s}\in\mathbb{R}^{N\times1}} \|{\bm s}\|_{1},~{\rm s.t}.~{\bm y}={\bf \Phi}{\bm h}={\bm \Phi}{\bf \Psi}^{-1}{\bm s}
\eeq
where ${\bm s}$ is the sparse representation of the original signal such that
\beq\label{spv}
{\bm s}={\bf \Psi}{\bm h},
\eeq
and ${\bf \Psi}\in\mathbb{C}^{N\times N}$ is a {\it representation} matrix. Linear programming can be used to find the sparse signal ${\bm s}$ for given ${\bm y}$, ${\bm \Phi}$, and ${\bf \Psi}$ in (\ref{CS}). Once $\widetilde{\bm s}$ is obtained from (\ref{CS}), ${\bm h}$ can be recovered as
\beq\label{est}
\widetilde{\bm h}={\bf \Psi}^{-1}\widetilde{\bm s}.
\eeq

A robust uncertainty principle in \cite{CaRoTa06} states that the number of minimum required observation samples,  {$M$}, for the perfect recovery of ${\bm s}$ in (\ref{CS}) is reduced as the ${\bm s}$ becomes more sparse. Concretely, for the perfect recovery of $K$-sparse ${\bm s}$ that includes $K$-nonzero and $(N-K)$-zero elements, the number of observation samples  {$M$} should  {fulfill}
%\beq \nonumber
$M\geq cK\ln{N},
$ %\eeq
where $c>0$ a small constant. Herein, note that the sparsity $K$ depends on the representation matrix ${\bf \Psi}$ in (\ref{spv}).

\subsection{Interpretation of channel feedback from a CS Perspective}
When we consider CS for channel feedback in communications, along with the recovery performance, we have to consider the feedback amount, which is an overhead in communications. Therefore, contrary to the CS in data processing, in which the original ${\bm h}$ is recovered from the observation  {${\bm y}\in\mathbb{C}^{M\times1}$}, the channel feedback in communications feeds back the $K$-sparse signal ${\bm s}\in\mathbb{C}^{K\times1}$ to recover ${\bm h}$ because  {$K<M$}, i.e.,

\begin{framed}
%\begin{itemize}
%\item
\vspace{-.2cm}
\noindent CS: ${\bm y}$ ( {$M$} observations) $\overset{\ell_1~{\text{minimization}}}{\xRightarrow{\hspace*{1.5cm}}}{\widetilde{\bm h}}$ ($N$ data recovery)\vspace{0.2cm}\\
%\item
channel feedback: ${\bm s}$ ($K$-sparse repr.) $\overset{{\text{$M$ feedback}}}{\xRightarrow{\hspace*{1cm}}}{\widetilde{\bm h}}$ ($N$ channel recovery)
%\end{itemize}
%\vspace{-.5cm}
\end{framed}

A feedback amount is assumed to be fixed to avoid additional overhead to inform it. Hence, regardless of the sparsity of ${\bm s}$, the Rx feeds back $M$ samples from ${\bm s}$ by using a selection matrix ${\bm S}\in\mathbb{R}^{M\times N}$ that is a binary matrix to select the most significant feedback information from ${\bm s}$. Each row of ${\bm S}$ selects one sparse channel. Precisely, the $n$th element is `$1$' and other elements in the row are `$0$'s if the $n$th element of ${\bm s}$ is selected to be fed back. Since each element of ${\bm s}$ could be selected at most once, each column of ${\bm S}$ includes at most one `$1$'.

{The} Rx feeds back the selected sparse CSI vector ${\bm s}'$ to a Tx, where ${\bm s}'$ is written as
%\beq\label{fb-info}
${\bm s}'={\bm S}{\bm s}\in\mathbb{C}^{M\times 1},
$ %\eeq
and a Tx recovers the channels from ${\bm s}'$ based on (\ref{spv}) and (\ref{est}) as follows:
\beq\label{CSr}
\widetilde{\bm h}={\bf \Psi}^{-1}{\bm S}^{\dag}{\bm s}' =
{\bf \Psi}^{-1}{\bm S}^{\dag}{\bm S}{\bm s}={\bf \Psi}^{-1}{\bm S}^{\dag}{\bm S}{\bf \Psi}{\bm h}.
\eeq

In (\ref{CSr}), if $M\geq K$, we can design ${\bm S}$ such that ${\bm S}^{\dag}{\bm S}{\bm s}=\overline{{\bm I}_{M}}{\bm s} = {\bm s}$, where $\overline{{\bm I}_{M}}$ is a diagonal matrix whose diagonal element is either $1$ or $0$, and $K$ of $M$ non-zero diagonal elements correspond to the non-zero elements of ${\bm s}$. As consequence, the Tx can recover ${\bm h}$ perfectly. Note that, contrary to the CS, a measurement matrix ${\bf \Phi}$ and an $\ell_1$ minimization are not required for channel feedback and the recovery, and that the design of ${\bf \Psi}$ and ${\bm S}$ is a critical part affecting the channel recovery performance in channel feedback.

\subsubsection{Design of ${\bf \Psi}$}
 The main purpose of a representation matrix ${\bf \Psi}$ is to transform ${\bm h}$ as sparse as possible, so that $K$ and also $M$ can be reduced without any loss of information. To this end, the representation matrix ${\bf \Psi}$ can be designed based on the {\it channel characteristics}. For example, there are various well-known transforms, such as the discrete Fourier/sine/cosine/Hartley transform (DFT/DST/DCT/DHT) and KLT, depending on the channel characteristics. If ${\bm h}$ itself is sufficiently sparse, we can set ${\bf \Psi}$ to an identity matrix ${\bm I}_{N}$. If the ${\bm h}$ is quasi-static (looks like a step function), a difference matrix ${\rm Tz}[[1,-1,0,\cdots,0]^T]$ will {best sparsify ${\bm h}$}, where ${\rm Tz}[{\bm a}]$ generates a Toeplitz matrix with ${\bm a}$ as its first row vector. A DFT matrix can be used to capture the frequency-domain correlated channels. The DST/DCT has good energy compaction property; thus, it can achieve near-optimal compression performance. However, DST/DCT does not perform very well if the channels are highly correlated \cite{LiCh93}. As mentioned in Introduction, a PCA using KLT is an optimal transform that can decorrelate the channels into a representation with the most sparse, non-redundant channels. In Section \ref{Sec:SCF}, we design the KLT matrix from channel statistics, such as spatial correlation, variance, and delay profile of ${\bm h}$.

\subsubsection{Design of ${\bm S}$}
After transforming the original channel ${\bm h}$ to ${\bm s}$, the selection matrix ${\bm S}$ can be designed according to the {\it sparsity characteristics} of ${\bm s}$. For example, if the sparse channels are distributed randomly over the sparse domain, ${\bm S}$ is designed to select the significant, sparse channels. In the case, along with the sparse channel values, the Rx needs to feed back the index of the selected channels, i.e., selection matrix ${\bm S}$ is a variable, which requires $\log_2 \prod_{m=0}^{M-1}(N-m)$ additional feedback bits per dimension to inform the index of `$1$' of each of $M$ rows. On the other hand, if the significant channels are typically located within a fixed sparse-domain regime, i.e., ${\bm s}$ is structured sparsity, ${\bm S}$ can be fixed to be implemented at both Tx and Rx.

\section{Proposed Sparse CSI Feedback}\label{Sec:SCF}
Based on PCA, we have a representation matrix ${\bf \Psi}$ that consists of eigenvectors of channel covariance matrix as \cite{Jo86Book,HuLi98,JoNeRiSe10}
\beq\label{KLT3}
\overline{\bf \Psi} = {\rm eig}\left( {\bm C}_{\bm h}\triangleq\E\left({\bm h}{\bm h}^H\right)\right)\in\mathbb{C}^{N\times N},
\eeq
where ${\rm eig}(\bm A)$ takes ${\bm U}^H$ from an eigenvalue decomposition such that ${\bm A}={\bm U}{\bm D}{\bm U}^H$. Here, ${\bm D}$ is a diagonal matrix whose diagonal elements are the eigenvalues of ${\bm A}$, and ${\bm U}$'s column vectors are the eigenvectors of ${\bm A}$. The PCA designs a selection matrix ${\bm S}_{\rm PCA}$ such that it selects the $M$ largest eigen values of $\overline{\bf \Psi}$. Hence, the most significant sparse channels will be selected. In a practical communications system, however, it is challenging to obtain the exact ${\bm C}_{\bm h}$, which motivates us to represent the channel covariance matrix ${\bm C}_{\bm h}$ with respect to the tractable channel statistics of ${\bm h}$.

To derive ${\bm C}_{\bm h}$ analytically, following Property \ref{prop3} regarding a CSI structure is useful.
%\begin{framed}
\begin{property}\label{prop3}
The CSI structure of ${\bm h}$ does not affect the CSI recovery performance in an SCF scheme. In other words, a restructured CSI with an arbitrary permutation matrix ${\bm P}\in\mathbb{R}^{N\times N}$, i.e., ${\bm P}{\bm h}$, provides the same recovery performance as the original ${\bm h}$ in an SCF scheme.
\begin{IEEEproof}
See Appendix \ref{Appendix:C}.
\end{IEEEproof}
\end{property}
%\end{framed}
Using the Property \ref{prop3}, without loss of generality, we structure a {\it spatial-and-frequency} domain channel vector ${\bm h}$ as follows:
\beq\label{ch}
{\bm h} = \vec \left(\left[ {\bm h}_1 \cdots {\bm h}_{N_f} \right] \right)=[{\bm h}_1^T \cdots {\bm h}_{N_f}^T ]^T \in\mathbb{C}^{N\times 1},
\eeq
where $\vec({\bm A})$ denotes a vectorization of a $m$-by-$n$ matrix ${\bm A}$ to form the $mn$-by-$1$ column vector obtained by stacking the columns of the matrix ${\bm A}$ on top of one another; $N_f$ is the number of subbands (subcarriers); ${\bm h}_n$ is the spatial-domain channel vector for frequency band $n$ that is modeled as
\beq\nonumber
{\bm h}_n=\vec\left({\bm H}(n)\right)\in\mathbb{C}^{N_rN_t \times 1},~n\in\{1,\ldots,N_f\};
\eeq
${\bm H}(n)\in\mathbb{C}^{N_r\times N_t}$ is the spatially correlated MIMO channel matrix of subcarrier $n$; $N_r$ and $N_t$ are number of receive and transmit antennas; and $N=N_r N_t N_f$. The spatially correlated MIMO channel is represented by \cite{KeScPeMoFr02}
\beq\nonumber%
{\bm H}(n)={\bm R}_{\rm r}^{\frac{1}{2}}{\bm H}_{\rm iid}(n)\left({\bm R}_{\rm t}^{\frac{1}{2}}\right)^H,
\eeq
where ${\bm R}_{\rm r}\in\mathbb{R}^{N_r\times N_r}$ and ${\bm R}_{\rm t}\in\mathbb{R}^{N_t\times N_t}$ are receive- and transmit-antenna correlation matrices, respectively (for a spatial correlation model of 2-dimension antenna, see Appendix \ref{Appendix:D}), and ${\bm H}_{\rm iid}(n)\in\mathbb{C}^{N_r\times N_t}$ is the uncorrelated, spatial-domain MIMO channel matrix of subcarrier $n$. The $(i,j)$th elements of ${\bm H}_{\rm iid}(n)$ represents a channel gain consisting of the path loss and the small scale fading between transmit antenna $j$ and receive antenna $i$. The channel elements are assumed to obey the complex normal distribution with a zero mean and a $\sigma_h^2$ variance, i.e., $\mathcal{CN}(0, \sigma_h^2)$, and be independent and identically distributed (i.i.d.).

The channel structure in (\ref{ch}) allows us to derive the closed form of ${\bm C}_{\bm h}$ as shown in Property \ref{prop1}.
%\begin{framed}
\begin{property}\label{prop1}
For the channel vector ${\bm h}$, whose structure follows (\ref{ch}), its covariance matrix is derived formally as
\beq\nonumber
{\bm C}_{\bm h} = {\bm C}_{\rm f} \otimes\left({\bm R}_{\rm t} \otimes {\bm R}_{\rm r}\right),
\eeq
where $\otimes$ represents Kronecker product of two matrices; ${\bm C}_{\rm f}={\rm Tz}[c_{1}^2,\cdots,c_{N_f}^2]$ is a frequency-domain covariance matrix; and $c_{n}^2$ is the correlation factor between the frequency domain channels ${\bm H}_{\rm iid}(1)$ and ${\bm H}_{\rm iid}(n)$. For the frequency domain channels generated by DFT of $L$-tap time domain channels with a delay profile ${\bm d}\in\mathbb{R}^{L\times1}$, the correlation factor $c_n^2$ is expressed as
\beq\nonumber
c_n^2 = \sigma_h^2\tr\left({\bm f}_1^{{\rm r},H}{\bm f}_n^{\rm r}\diag({\bm d})\right),~\forall n\in\mathcal{N}=\{1,\cdots,N_f\}.
\eeq
Here, ${\bm f}_n^{\rm r}\in\mathbb{C}^{1\times L}$ is the $n$th row vector of ${\bm F}_L\in\mathbb{C}^{N\times L}$ that consists of the first $L$ column vectors of $N$-point DFT matrix ${\bm F}\in\mathbb{C}^{N\times N}$.
\begin{IEEEproof}
See Appendix \ref{Appendix:A} for the proof.
\end{IEEEproof}
\end{property}
%\end{framed}

\begin{algorithm}[h]
\caption{: Sparse CSI Feedback (SCF)}\label{alg:SCF}
%\algsetup{indent=-10em,linenosize=\footnotesize,linenodelimiter=.}
\begin{algorithmic}
\STATE{\begin{enumerate}
\item {\bf Offline/Online Mode}: Setup
\begin{enumerate}
\item measure the required channel statistics, namely $\sigma_h^2$, ${\bm d}$, ${\bm R}_{\rm t}$, and ${\bm R}_{\rm r}$.
\item compute ${\bm C}_{\bm h}$ from Property \ref{prop1}.
\item using ${\bm C}_{\bm h}$, compute $\overline{\bm \Psi}$ in (\ref{KLT3}).
\item store $\overline{\bm \Psi}$ at both Tx and Rx.
%\item set the selection matrix ${\bm S}=[{\bm I}_M~ {\bm 0}_{M\times(N-M)}]$, where ${\bm I}_M$ is an $M$-dimensional identity matrix and ${\bm 0}_{M\times(N-M)}$ is an $M$-by-$(N-M)$ zero matrix.
\end{enumerate}
\item {\bf Rx's}: Feedback from each Rx, ${\bm S}={\bm S}_{\rm KLT}(M)$
\begin{enumerate}
\item estimate channels ${\bm H}(n)$ for all $n=\{1,\ldots,N_f\}$.
\item construct ${\bm h}$ in (\ref{ch}).
\item get a sparse channel representation ${\bm s}=\overline{\bm \Psi}{\bm h}$.
\item generate a selected sparse vectors ${\bm s}'={\bm S} {\bm s}$.
\item feed back ${\bm s}'$, and ${\bm S}$ if it is needed.
\end{enumerate}
\item {\bf Tx}: Recovery at Tx from (\ref{CSr})
\begin{enumerate}
\item recover the channels as $\widetilde{\bm h} = \overline{\bm \Psi}^{-1}{\bm S}^{\dag}{\bm s}'$.
\end{enumerate}
\end{enumerate}}
\end{algorithmic}
\end{algorithm}

\begin{remark}
The representation matrix $\overline{\bm \Psi}$ is fixed at both Tx and Rx and no additional eigen decomposition and feedback are required, especially, when the Rx is nomadic and thus the covariance matrix ${\bm C}_{\bm h}$ is static \cite{AdNaAhCa13}. Only the $M$-selected sparse channels, ${\bm s}'$, need to be fed back for CSI recovery at a Tx.
\end{remark}

\begin{remark}
The time variation of channel statistics is caused by the movement of a Rx, hence the offline estimates may be outdated and are needed to be updated by feedback. Depending on Rx mobility, each Rx measures the channel variance delay profile and feeds back them to a Tx, so that the Tx can update $\overline{\bm \Psi}$. The sporadic update of the statistics can dramatically reduce the feedback information compared to the update of $\overline{\bm \Psi}$ itself, and it alleviates the high overhead feedback and Rx complexity issues.
\end{remark}

\begin{remark}
One alternative implementation of ${\bm C}_{\bm h}$ in (\ref{KLT3}) is an empirical cumulative moving average as $\widetilde{{\bm C}_{\bm h}}\triangleq\frac{1}{t}\!\!\sum_{t'=1}^{t'=t} \left\{{\bm h}_{t'}{\bm h}_{t'}^H\right\}\in\mathbb{C}^{N\times N}$, where $t$ is update time \cite{GwKuVl12}. By updating every $T$ interval, feedback amount can be reduced. However, after $\widetilde{{\bm C}_{\bm h}}$ is numerically evaluated at an Rx, the new KLT matrix $\widetilde{\bf \Psi}$ should be updated at both Tx and Rx. Since the Rx has to compute the $\widetilde{{\bm C}_{\bm h}}$ and recalculate $\widetilde{\bf \Psi}$, the computational complexity may arise as an issue at the Rx that has insufficient computing capability. Furthermore, the update of $\widetilde{\bf \Psi}$ still requires channel feedback overhead.
\end{remark}

\section{Compressive Feedback Error Analysis}\label{Sec:Error-Analysis}
The quality of the channel recovery depends on the level of compression. The channel recovery error is expected to be more severe when the compression ratio $\gamma(M)$ is high, while it decreases as $\gamma(M)$ decreases. In practical systems, it is often necessary to give a performance guarantee to the users. In such cases, a quantitative analysis on the tradeoff between the compression ratio and the channel recovery performance is useful, since it specifies the constraint on how much compression can be tolerated for a given QoS requirement.

Following the discussion in Section \ref{Sec:CF-CS}, the compressive feedback error is considered solely from the compression, and does not include the estimation and quantization errors. The NMSE in (\ref{NMSE}) is calculated using ${\bm h}$ in (\ref{ch}) and $\widetilde{\bm h}$ in Algorithm \ref{alg:SCF}. In the proposed SCF using KLT, the amount of feedback (and correspondingly the compression ratio) is determined by the number of non-zero elements in the selection matrix ${\bm S}$. Without compression ($\gamma(M) = 1$), the amount of feedback would be equal to $N = N_r N_t N_f$ (the dimension of ${\bm h}$). In reality, due to correlation in frequency and spatial domain of the antennas, the actual number of dimension occupied by ${\bm h}$ is usually less than $N$. Considering that any realization of ${\bm h}$ can be expressed as ${\bm h} = {\bm C}_{\bm h}^{\text{\textonehalf}} {\bm a}$ where ${\bm a}\sim\mathcal{C}(0,{\bm I}_N)$, the number of non-zero elements $N'$ in ${\bm a}$ required to fully represent ${\bm h}$ is the same as the rank of ${\bm C}_{\bm h}$, i.e., $\rank({\bm C}_{\bm h})$. From the fact that we can achieve $\delta(M) = 0$ when $N' \geq\rank({\bm C}_{\bm h})$, a maximum distortion-free compression ratio denoted by $\gamma^*$ is derived as
\beq
\label{gamma-s}
\gamma^{*} = \frac{ N }{\rank( {\bm C}_{\bm h} )} \\
=\underset{\triangleq\;\gamma_{ f}}{\underbrace{\frac{N_f}{\rank( {\bm R}_{\rm f} )}}} \underset{\triangleq\;\gamma_{ t}}{\underbrace{\frac{N_t}{\rank( {\bm R}_{\rm t} )}}}
\underset{\triangleq\;\gamma_{ r}}{\underbrace{\frac{N_r}{\rank({\bm R}_{\rm r})}}}
\triangleq \gamma_{ f}\gamma_{ t}\gamma_{ r}
% \underset{\triangleq\;\gamma_f}{\underbrace{\frac{N_f}{\rank( {{\rm Tz}[c_{1}^2,\cdots,c_{N_f}^2]} )}}}.
\eeq
The result in (\ref{gamma-s}) is obtained by choosing the selection matrix ${\bm S}$ to extract components of ${\rm eig}({\bm C}_{\bm h})$ corresponding to the non-zero $M=N'$ eigenvalues and by using Property \ref{prop1}. %Setting $\gamma > \gamma^*$ will not be useful in this case, as the additional feedback does not carry any useful information.
An interesting remark from (\ref{gamma-s}) is as follows.
\begin{remark}
The maximum distortion-free compression ratio $\gamma^*$ is the product of the individual distortion-less compression ratio at each of the domains, i.e., $\gamma^*=\gamma_{ f}\gamma_{ t}\gamma_{ r}$ where $\gamma_{ f}$, $\gamma_{t}$, and $\gamma_{ r}$ are at the frequencies, transmit antennas, and receive antennas, respectively. Therefore, the effect of low rank CMs and the corresponding magnitude of the distortion-less compression ratios will be more pronounced when the correlation is present in multiple domains due to the multiplying effect as concretely shown in (\ref{gamma-s}).
\end{remark}

When higher compression ratio is desired, some of the components of ${\bm s}$ in (\ref{CSr}) corresponding to the non-zero eigenvalues have to be discarded as well, resulting in the recovery distortion. As described in the earlier section, following the idea of PCA, the best selection strategy is to discard the elements that correspond to the smallest eigenvalue first. Denoting the number of principle components that are kept as $M < N'$, the NMSE $\delta(M)$ is derived as (see Appendix \ref{Appendix:E})
\beq\label{eqn:deltaM}
\delta(M) = \frac{\sum\text{uncaptured $(N'-M)$ principal components}}{\sum\text{all $N'$ principal components}}.
\eeq
The NMSE $\delta(M)$ can be interpreted as the {\it best possible distortion} for a given compression ratio of $\gamma={N}/{M}$, and it is also known as a {\it distortion of data recovery} in the CS context \cite{CaRoTa06}. In other interpretation, the minimum number of principal components to be kept for a given compression ratio $\gamma$ is given by $M = {N}/{\gamma}$, and the selection matrix ${\bm S}$ will contain $M$ non-zero components. The resulting NMSE $\delta(M)$ can then be calculated using (\ref{eqn:deltaM}). We can also use (\ref{eqn:deltaM}) in system design when allocating the feedback bandwidth for a given QoS constraint, e.g., NMSE and bit-error-rate (BER).

\begin{figure}[!t]
\psfrag{a                                                          }[lc][cc][.8][0]{\sf SCF: Monte Carlo}
\psfrag{c}[lc][cc][.8][0]{\sf SCF: Analysis $f(\mu)$}
\psfrag{e}[lc][cc][.8][0]{\sf SCF: Analysis $\overline{f}$ in (\ref{BERanalysis})}
\psfrag{x}[cc][cc][.8][0]{\sf feedback compression ratio, $\gamma_{\rm fb}$ (refer to Section \ref{Sec:Sim}.C)}
\psfrag{y}[cc][cc][.8][0]{\sf BER}
\begin{center}
\epsfxsize=0.43\textwidth \leavevmode\epsffile{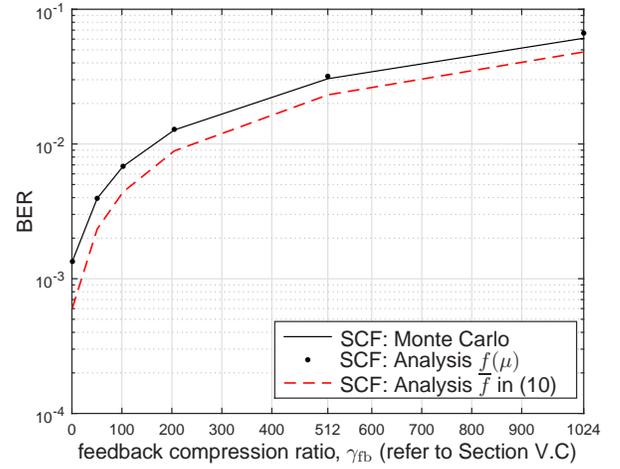}
\caption{BER evaluation and BER analysis justification.}% performance evaluation and with $N_{th}=64$, $N_r=1$, $N_f=64$, 16-QAM, and $EsNo=-20\dB$ for a single user.}
% BER_analysis2_fixed_SNR.m
\label{Fig:0}
\end{center}
\end{figure}

Now, we analyze the system performance in terms of the BER to get the answer how the choice of compression ratio $\gamma(M)$ affects the BER performance. Assuming that beamforming is used at both the spatial and frequency dimension, and using the fact that the subspace of the channel estimation error and that of the channel estimate are orthogonal to one another, we can derive an effective signal-to-noise ratio, denoted by $\mu$, as $\mu=\|\widetilde{\bm h}\|^2 / \sigma^2$, where $\widetilde{\bm h}$ is the reconstructed channel vector and $\sigma^2$ is the noise variance. With $\mu$, the BER can then be derived as a convex function of $\mu$; for example, $f(\mu) = {3}/{4} \mathcal{Q}(\sqrt{{\mu}/{5}} ) + {1}/{2} \mathcal{Q}(\!\sqrt{{3\mu}/{5}} ) \!- {1}/{4} \mathcal{Q}(\!\sqrt{\mu} )$ for $16$-QAM modulation with Gray bit mapping \cite{PrSa07book}, where $\mathcal{Q}(x) = \int_x^\infty (2\pi)^{-0.5} \exp(-0.5 t^2) \text{d}t$ is a standard Q-function. By the convexity of the Q-function and invoking Jensen's inequality, the lower bound on the average BER $\overline{f}$ can then be obtained as follows (refer to the notation in Appendix \ref{Appendix:E}):
\beq
\begin{split}\label{BERanalysis}
\overline{f} &
\triangleq \E_{\widetilde{\bm h}}\left(f(\mu)\right) \geq f(\E[\mu])=
f\left(\sigma^{-2} {\rm tr}\left(\E\left( \widetilde{\bm h} \widetilde{\bm h}^H \right) \right)\right)\\
&=f\left( \sigma^{-2} {\rm tr}\left({\bm C}_{\bm h}^{\frac{1}{2}} {\bm S}^\dag {\bm S} \E\big({\bm a} {\bm a}^{H} \big) ({\bm S}^\dag {\bm S})^{H} {\bm C}_{\bm h}^{\frac{1}{2}}\right) \right) \\
&= f\left( \sigma^{-2}{{\rm tr}\left({\bm S} {\bm D} \right)} \right)=
f\left( \sigma^{-2}{{\rm tr}\left({\bm D} \right)\left(1-\delta(M)\right)} \right).
\end{split}
\eeq
From (\ref{BERanalysis}), we can clearly see the relationship between the effect of compression ratio $\delta(M)$ on the BER performance $\overline{f}$.

For the verification of the BER and NMSE analyses, please refer to Fig. \ref{Fig:0} and Fig. \ref{Fig:2}(a). The BER ($16$-QAM) is evaluated for a single user using beamforming $\tilde{\bm h}^{H}$, when $N_{t,h}=64$, $N_{t,v}=1$, $N_r=1$, and $N_f=64$, i.e., $N=2^{12}=4096$. %, and $EsNo=-20\dB$.

\begin{table*}[!t]
\centering
\caption{\vspace{.2cm}CSI Feedback (FB) Schemes with $\gamma=N/M$ and $Q$-Bit Quantization.}
\label{Tab:2}
\resizebox{.85\textwidth }{!}{
\begin{tabular}{c|c|c|c|c|c|c }\hline
Schemes  &  ${\bm \Psi}$& ${\bm S}$ & FB info. & Total number of FB bits & CSI structure & Acronym  \\\hline
\multirow{2}*[0.05cm]{Freq-domain CSI FB}  & \multirow{2}*[0.05cm]{${\bm I}_{N}$} & \multirow{2}*[0.05cm]{${\bm S}_{\rm FCF}(M)$} & \multirow{2}*[0.05cm]{${\bm s}'$} & \multirow{2}*[0.05cm]{$2MQ$}& ${\bm h}$ in (\ref{ch})  & FCF-f1 \\\cline{6-7}
   & &  &  &  & ${\bm h}'$ in (\ref{ch2}) & FCF-f2\\\hline
\multirow{4}*[0.05cm]{Time-domain CSI FB} & \multirow{4}*[0.05cm]{${\bm F}^{-1}$}&\multirow{2}*[0.05cm]{${\bm S}_{\rm TCF}(M)$}  & \multirow{2}*[0.05cm]{${\bm s}'$} & \multirow{2}*[0.05cm]{$2MQ$} & ${\bm h}$ in (\ref{ch})&  TCF-f1\\\cline{6-7}
 &  &  &  &   & ${\bm h}'$ in (\ref{ch2})&  TCF-f2\\\cline{3-7}
 &  & \multirow{2}*[0.05cm]{variable} &\multirow{2}*[0.05cm]{${\bm S}$ and ${\bm s}'$} & \multirow{2}*[0.05cm]{$2MQ+2 \log_2 \prod_{m=0}^{M-1}(N-m)$} &${\bm h}$  in (\ref{ch})&{TCF-v1}\\\cline{6-7}
 & &  &  &  & ${\bm h}'$ in (\ref{ch2})& TCF-v2\\\hline
\multirow{2}*[0.05cm]{Sparse-domain CSI FB}  & \multirow{2}*[0.05cm]{$\overline{\bm \Psi}$} & ${\bm S}_{\rm PCA}(M)$ & ${\bm s}'$ & $2MQ$ & \multirow{2}*[0.05cm]{${\bm h}$ in (\ref{ch})}& SCF-f \\\cline{3-5}\cline{7-7}
   & & variable & ${\bm S}$ and ${\bm s}'$ & $2MQ+2\log_2 \prod_{m=0}^{M-1}(N-m)$& & SCF-v\\\hline
Full CSI FB  & --& --  & ${\bm h}$ & $2NQ$ (no compression) & --& Full channel feedback\\\hline
\end{tabular}
}
\end{table*}

\section{Performance Evaluation and Discussion}\label{Sec:Sim}
With the analytical framework proposed in the earlier sections, we further verify the proposed SCF method by comparing it with other channel feedback schemes summarized in Table \ref{Tab:2}. We consider OFDM system and compare basically three schemes, namely, frequency-domain channel feedback, time-domain channel feedback, and the proposed SCF using $\overline{\bf \Psi}$, which are denoted by {FCF}, {TCF}, and {SCF}, respectively, with the suffixes `{\rm f}' and `{\rm v}' for fixed and variable ${\bm S}$, respectively. To clearly compare the performance, all comparison results are obtained for fixed $\sigma_h^2$ and ${\bf d}$. After briefly introducing the FCF and TCF, we show the comparison results.  For the fixed ${\bm S}_{\rm PCA}$ we fix it by ${\bm S}_{\rm PCA}(M)\triangleq [{\bm I}_M ~{\bm 0}_{M,N-M+1}]$.

\subsection{Frequency-domain CSI Feedback (FCF)}
For the sake of comparison, an identity matrix ${\bm I}_{N}$ is employed as a representation matrix. The identity representation matrix gives us the simplest way that reduces the feedback information by feeding back the partially, directly selected information from the original channel ${\bm h}$ as follows:
\beq\label{FCFch}
{\bm s}' = {\bm S}_{\rm FCF}{\bm s}={\bm S}_{\rm FCF}{\bm I}_N{\bm h}={\bm S}_{\rm FCF}{\bm h},
\eeq
where ${\bm S}_{\rm FCF}$ is a selection matrix. Since the original channel ${\bm h}$ is the aggregation of frequency-domain spatial channels, we call this scheme as a FCF methods. Having the knowledge of predetermined ${\bm S}_{\rm FCF}$ and feedback information ${\bm s}'$, the Tx can obtain the estimate of the original channels as follows:
\beq
\widetilde{\bm h} = {\rm{intp}}\left({\bm S}_{\rm FCF}^{\dag}{\bm s}'\right),
\eeq
where ${\rm{intp}}(\cdot)$ represents an interpolation function to recover the unselected channels in (\ref{FCFch}). The recovery performance of FCF depends on the interpolation method, the selection matrix, and the original channel structure\footnote{Contrary of the SCF method, in which the channel structure does not affect the CSI recovery performance as shown in Property \ref{prop3}, the channel structure generally affects on the recovery performance.}. Designing the interpolator and selection matrix is out of the scope of this work. For the channel structure, we consider another structure defined as
\beq\label{ch2}
{\bm h}'= \vec\left(\left[{\bm h}_1\cdots {\bm h}_{N_f}\right]^T\right),
\eeq
which is similar to the bundled channel structure in \cite{MaHaKo13}.

In simulation, we fix the selection matrix of FCF by ${\bm S}_{\rm FCF}$ that selects the channels located in equidistance of frequency domain axis to feed back. For the interpolation, we employ a spline interpolation method \cite{Bo78Book}. As mentioned, since an optimal selection matrix depends on the interpolation method and channel distribution, we do not consider a variable selection matrix for the FCF. Though the FCF method requires low computational complexity at the transceiver, the compression ratio $\gamma$ is generally desired to be low to achieve reliable CSI recovery performance in communications.

\subsection{Time-domain CSI Feedback (TCF)}
An inverse DFT (IDFT) matrix ${\bm F}^{-1}$ is employed as a representation matrix. Since the Rx feeds back the IDFT of the aggregation of frequency-and-spatial domain channels, for simple denotation, we address this scheme as a TCF method.

Fixed and variable selection matrices are considered for the TCF method. The fixed selection matrix is designed to capture the most significant channels from ${\bm s}$. In simulation, we fix the selection matrix as ${\bm S}_{\rm TCF}= \bmats {\bm I}_{M/2} & {\bm 0}_{M/2,N-M/2+1}\\ {\bm 0}_{M/2,N-M/2+1}& {\bm I}_{M/2} \emats$ to capture the most significant time-domain channels from ${\bm h}'$. The variable selection matrix selects the most significant $M$ elements in ${\bm s}$.

Compared to the FCF method, the CSI recovery of TDF does not require interpolation, yet it requires re-transformation, i.e., DFT, to recover the CSI at the Tx, which is  {the} same as the procedure in (\ref{CSr}). The recovery performance depends on the selection method and the sparsity of ${\bm s}$. Since the IDFT and DFT are already implemented in the OFDM transceiver, and thus the TCF method is natural to be considered for the channel feedback of OFDM systems.

\subsection{Simulation Results}
The number of feedback bits for real and imaginary values of each scheme is summarized at the fifth column in Table \ref{Tab:2} when the data compression ratio is fixed by $\gamma=N/M$ and $Q$-bit quantization is employed. Since the different schemes generally require different numbers of feedback bits, we define a {\it feedback compression ratio} as
%\begin{framed}
\beq\nonumber
\gamma_{\rm fb} = \frac{\text{Total number of feedback bits {\it without} compression}}{\text{Total number of feedback bits {\it with} compression}}
\eeq
%\end{framed}
and fairly compare the performances for the same $\gamma_{\rm fb}$. %\textcolor[rgb]{0.00,0.50,0.00}{In 
For the performance metric, we consider channel recovery performance at Tx and communications performance at Rx, namely an NMSE and a BER ($16$-QAM). We consider one Tx and four Rx's, i.e., an MU massive MIMO system. The Tx and each Rx have $64$ and two 2-D antennas with the configuration of $N_t=64$ ($N_{t,h}=N_{t,v}=8$) and $N_r=2$ ($N_{r,h}=2,~N_{r,v} =1$). The Tx is located at the center of $1{\rm km}$-by-$1{\rm km}$ square-shaped coverage. Four Rx's are uniformly located within the coverage in each channel realization, and the corresponding large-scale fading is set into the variance $\sigma_h^2$ of the uncorrelated Rayleigh fading channels accordingly. The path loss model follows that $-123+10\log_{10}(l^{-3.76})$, where $l$ is distance between Tx and Rx in kilometer. Channel delay follows an exponential decaying profile and the number of channel taps is seven, i.e., $L=7$. Tx and Rx antenna spatial correlation factors are $\rho_t=0.8$ and $\rho_r = 0.5$, respectively. %Note that Tx correlation is higher than Rx correlation due to the large number of Tx antennas in the limited space.
All Rx's share $64$ subcarriers, i.e., $N_f=64$, and feed back their own CSI, individually and independently. $12$ bits are used for the quantization of the feedback symbols, i.e., $Q=12$. The Tx supports multiple Rx's by using a zero-forcing-based MU-MIMO precoding, which can be obtained from the aggregated CSIs. System bandwidth is $10\MHz$ and maximum transmit power is $43\dBm$ and noise variance at the Rx is set to be $-174\dBm / \Hz$.

\begin{figure}[!t]
\psfrag{z                                             }[lc][cc][.8][0]{\sf FCF-f1} % added
\psfrag{b                 }[lc][cc][.8][0]{\sf FCF-f1} % added
\psfrag{c}[lc][cc][.8][0]{\sf FCF-f2}
\psfrag{e}[lc][cc][.8][0]{\sf TCF-f1}
\psfrag{n}[lc][cc][.8][0]{\sf TCF-f2}
\psfrag{o}[lc][cc][.8][0]{\sf TCF-v1}
\psfrag{s}[lc][cc][.8][0]{\sf TCF-v2}
\psfrag{u}[lc][cc][.8][0]{\sf SCF-f: simulation}
\psfrag{a}[lc][cc][.8][0]{\sf SCF-f}
\psfrag{r}[lc][cc][.8][0]{\sf SCF-f: analysis in (\ref{eqn:deltaM})}
\psfrag{v}[lc][cc][.8][0]{\sf SCF-v}
\psfrag{x}[cc][cc][.8][0]{\sf feedback compression ratio, $\gamma_{\rm fb}$}
\psfrag{y}[cc][cc][.8][0]{\sf NMSE}
\psfrag{m}[cc][cc][.8][0]{\sf BER}
\begin{center}
\subfigure[\!\!\!\!\!]{%
\epsfxsize=0.43\textwidth \leavevmode
\epsffile{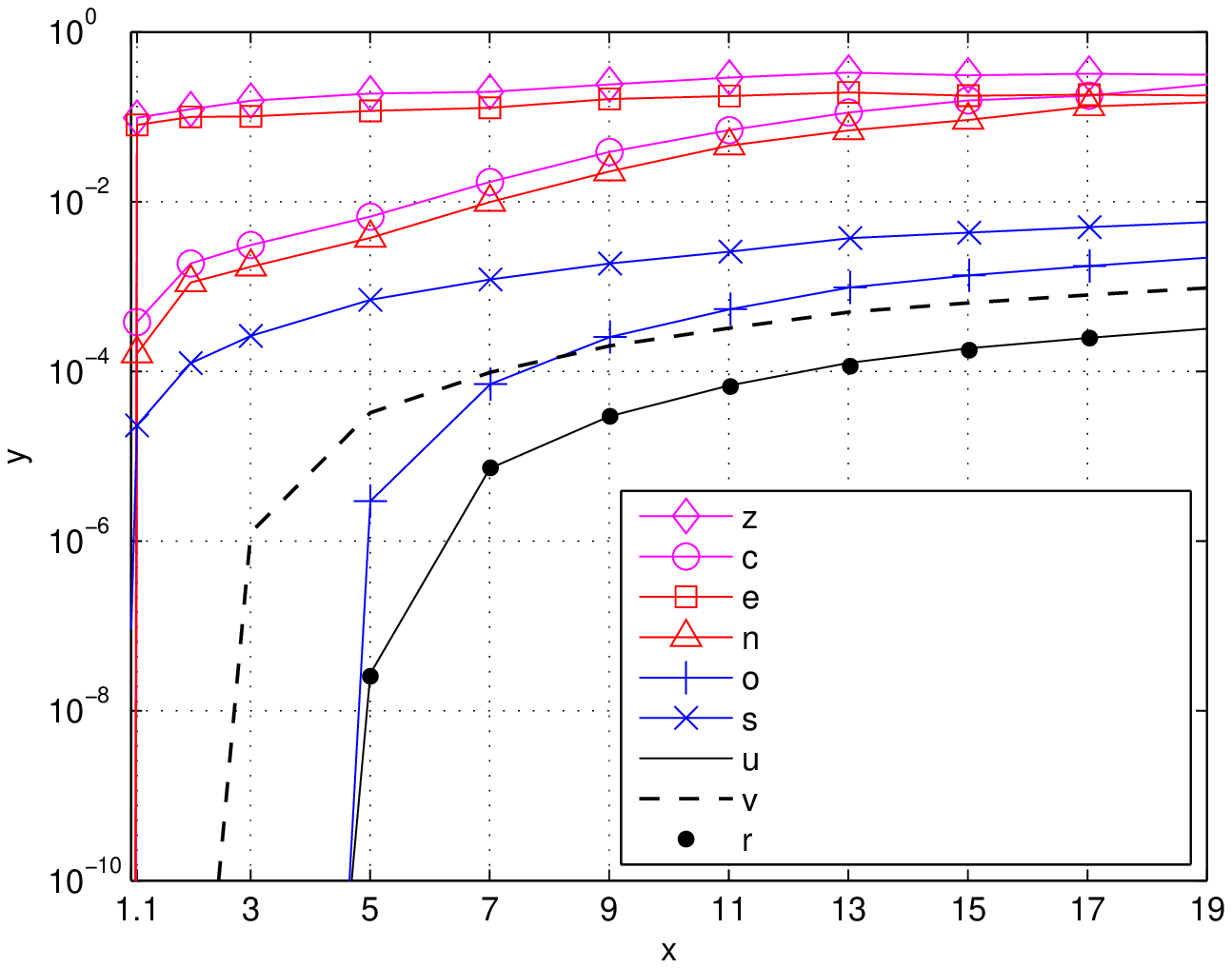}}\\
\subfigure[\!\!\!\!\!]{%
\epsfxsize=0.43\textwidth \leavevmode
\epsffile{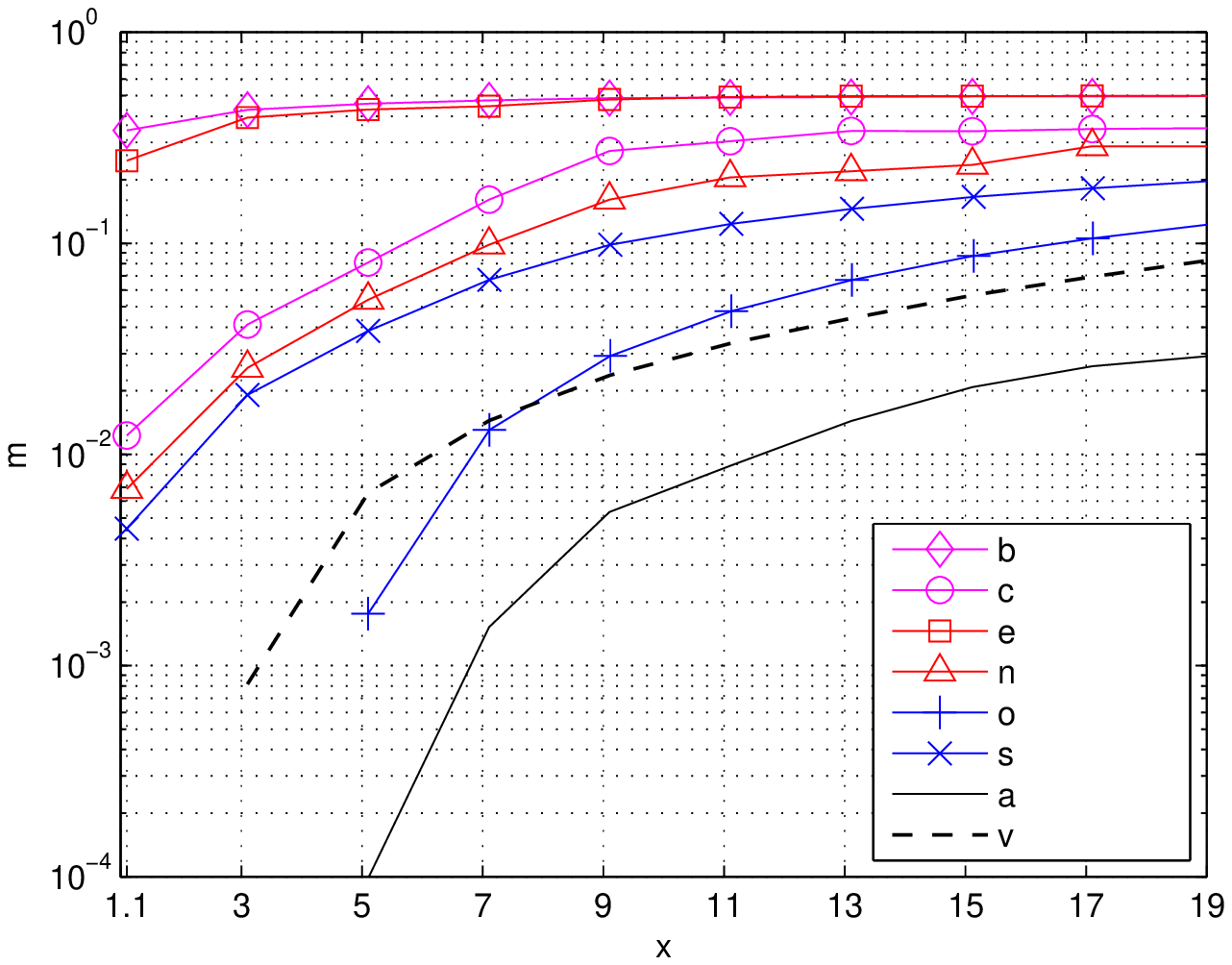}}
\caption{Performance comparison over feedback compression ratio $\gamma_{\rm fb}$. (a) NMSE at Tx. (b) BER at Rx.}
\label{Fig:2}
% BER_NMSE2.m
\end{center}
\end{figure}

In Figs. \ref{Fig:2}(a) and (b), the NMSE and BER (averaged over four Rx's) are shown, respectively. From the results, we see that the communication performance BER is directly affected by the CSI recovery performance, i.e., NMSE, because the MU interferences increases as the CSI uncertainty increases. We observe the agreement between the NMSE analysis in (\ref{eqn:deltaM}) and the numerical result. The channel recovery performance of FCF-f1 is very poor. This is because the variation of channel elements of ${\bm h}$ in (\ref{ch}) is significant, so that the interpolator cannot recover the original channels. The serious degradation of interpolation performance can be mitigated by structuring the original channels to ${\bm h}'$ in (\ref{ch2}), i.e., FCF-f2. On the other hand, the performance of TCF-f1 is also very poor as the channels after IDFT of ${\bm h}$ are distributed randomly over the time domain. Hence, the fixed selection matrix ${\bm S}_{\rm TCF}$ does not capture the significant channels, resulting in huge loss of information. Similar to FCF, restructuring ${\bm h}$ to ${\bm h}'$ can improve TCF performance as shown with TCF-f2. Herein, the significant time-domain channels are located most likely at the boundary of the time domain, which are well aligned to ${\bm S}_{\rm TCF}$. Further performance improvement can be achieved by selecting the channels according to their strength with allowing the additional feedback for the variable selection matrix, namely TCF-v1 and TCF-v2. From one interesting result that TCF-v1 outperforms TCF-v2, we see that the time-domain channel of ${\bm h}$ is more sparse than that of ${\bm h}'$. Contrary of TCF, in which the variable selection improves the performance effectively, for the proposed SCF, additional feedback for a variable selection matrix does not improve the performance. In other words, fortunately, the sparse-domain channels are mainly located within specific range that corresponds to the fixed selection matrix ${\bm S}_{\rm KLT}$. Since the channels' sparsity is high and their distribution are sufficiently captured by ${\bm S}_{\rm KLT}$ already, the additional feedback decreases the compression performance. Up to $\gamma_{\rm fb}=4$, no compression errors arise for {TCF-v1} and {SCF-f} and, thus, which is the same as the optimal BER with perfect CSI, i.e., a Full channel feedback. The compression performance improvement through the proposed SCF is significant. For example, to achieve $0.03$ BER performance, the {SCF-f} can reduce the feedback information around by half compared to the {TCF-v1}, i.e., from $\gamma_{\rm fb}=9$ to $\gamma_{\rm fb}=19$. The numerical results verify that the proposed {SCF-f} achieves always the best compression performance with the smallest NMSE and BER for given $\gamma_{\rm fb}$, i.e., for given feedback amount.

\begin{figure}[!t]
\psfrag{z                }[lc][cc][.8][0]{\sf FCF-f1} % added
\psfrag{c}[lc][cc][.8][0]{\sf FCF-f2}
\psfrag{e}[lc][cc][.8][0]{\sf TCF-f1}
\psfrag{n}[lc][cc][.8][0]{\sf TCF-f2}
\psfrag{o}[lc][cc][.8][0]{\sf TCF-v1}
\psfrag{s}[lc][cc][.8][0]{\sf TCF-v2}
\psfrag{u}[lc][cc][.8][0]{\sf SCF-f}
\psfrag{v}[lc][cc][.8][0]{\sf SCF-v}
\psfrag{x}[cc][cc][.8][0]{\sf Feedback amount reduction $\%$}
\psfrag{y}[cc][cc][.8][0]{\sf Spectral efficiency degradation $\%$}
\begin{center}
\epsfxsize=0.43\textwidth \leavevmode
\epsffile{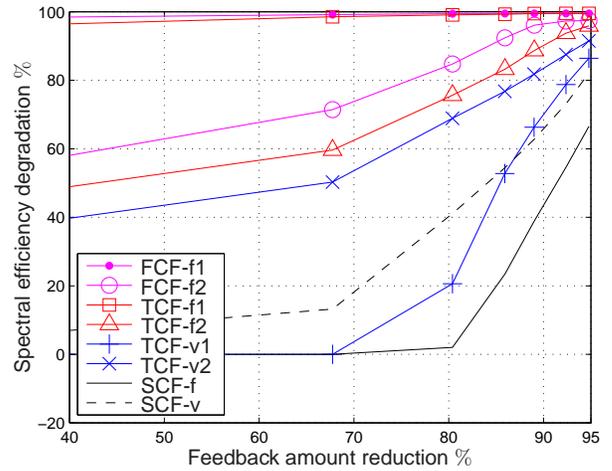}
\caption{Spectral efficiency reduction over feedback amount reduction.}
\label{Fig:3}
% Rate.m
\end{center}
\end{figure}

In Fig. \ref{Fig:3}, we evaluate spectral efficiency (SE) degradation from a SE bound, which is obtained from Full channel feedback. The SE is defined by the sum of each user's throughput $\log_2(1+{\rm SINR})$. The SE reduction of each scheme is shown over feedback amount reduction, i.e., $(\gamma_{\rm fb}-1)/\gamma_{\rm fb} \times 100 ~\%$. From the results, we can quantify how much the communication performance degrades due to the compression of each FB scheme. For example, TCF-v2 achieves SE that is degraded from its bound by $20\%$ with $80\%$ feedback amount reduction, yet the proposed SCF-f can achieve near optimal performance with feedback amount reduction up to $80\%$ (SE reduction by $2\%$ with feedback amount reduction by $80\%$). As shown, the SE of the proposed SCF-f achieves the best SE regardless of the feedback amount reduction. One interesting observation is that we can still communicate with $95\%$-reduced CSI information, even though the SE is degraded by about $65\%$. Such a reduced SE could be one possible application for low-rate transmission, e.g., control signal from data collection center to distributed multiple sensors in sensor networks.

\begin{figure}[!t]
\psfrag{a                   }[lc][cc][.7][0]{\sf SCF-f}
\psfrag{u}[lc][cc][.7][0]{\sf Full channel feedback}
\psfrag{e}[cc][cc][.8][0]{\sf $\gamma_{\rm fb}\!=\!4$}
\psfrag{f}[cc][cc][.8][0]{\sf $\gamma_{\rm fb}\!=\!2$}
\psfrag{g}[cc][cc][.8][0]{\sf $\gamma_{\rm fb}\!=\!1.5$}
\psfrag{h}[cc][cc][.8][0]{\sf $\gamma_{\rm fb}\!=\!1.2$}
\psfrag{c}[cc][cc][.8][0]{\sf $7^2\times 2$}
\psfrag{o}[cc][cc][.8][0]{\sf $6^2\times 2$}
\psfrag{n}[cc][cc][.8][0]{\sf $5^2\times 2$}
\psfrag{m}[lc][cc][.8][0]{\sf $4^2\times 2$}
\psfrag{v}[lc][cc][.8][0]{\sf $3^2\times 2$}
\psfrag{s}[rc][cc][.8][0]{\sf $N_t\times N_r=8^2\times 2$}
\psfrag{z}[rt][cc][.8][0]{\sf $\gamma_{\rm fb}=1$}
\psfrag{x}[cc][cc][.8][0]{\sf Amount of feedback, bytes/user/subcarrier}
\psfrag{y}[cc][cc][.8][0]{\sf Spectral efficiency, bits/sec/Hz}
\begin{center}
\epsfxsize=0.43\textwidth \leavevmode
\epsffile{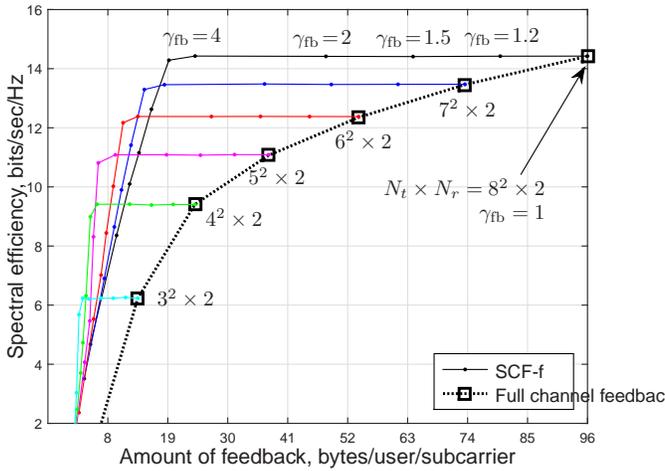}
\caption{Spectral efficiency comparison over feedback amount.}
\label{Fig:4}
\end{center}
% Rate_bytes.m (for origin), Rate_bytes_for_each.m (for each configuration)
\end{figure}

In Fig. \ref{Fig:4}, we evaluate the SE's of the proposed SCF-f scheme for various $N_t=a^2$ ($N_{t,v}=a$ and $N_{t,h}=a$) over the actual amount of feedback bytes. Square marks represent SEs with full channel feedback. As we increase a feedback compression ratio $\gamma_{\rm fb}$, the actual amount of feedback bytes (value in x-axis) decreases, while SE (value in y-axis) is retained up to a certain level of $\gamma_{\rm fb}$ and turn to decrease. The results verify that always the proposed SCF-f can reduce feedback amount without performance compromise. From the results, interestingly, we can observe that the maximum SE at given feedback amount is obtained not necessarily from larger number of $N_t$. For example, if the the feedback amount is limited by $8{\rm bytes/user/subcarrier}$ due to the uplink capacity, the best choice of $N_t$ is $25$ rather than $36$, $49$, and $64$. This particular observation provides important message to us that we have to consider the uplink capacity limitation to maximize downlink SE in the communications systems with feedback.

\section{Conclusion}\label{Sec:Conclusion}
We have considered a compression method to feed back CSI for large-scale MIMO systems. A covariance matrix of spatially correlated Rayleigh fading channels has been analytically modeled and used to sparsify the original CSI based on PCA. From intensive performance evaluation of NMSE, BER, and SE, we have justified that the proposed sparse CSI feedback method can reduce the CSI amount significantly and effectively.

\appendices

\renewcommand{\theequation}{\thesection.\arabic{equation}}

\setcounter{equation}{0}
\section{2-Dimensional Spatial Correlation}\label{Appendix:D}
\newcounter{mytempeqncnt}
\begin{figure*}[b]
\hrulefill%
\vspace*{4pt}
\normalsize
\setcounter{mytempeqncnt}{\value{equation}} %
\setcounter{equation}{0}%
\beq
\begin{split}\label{Tm}
{\bm T}_m &= {\rm Tz}\left[\rho_{t}^{\sqrt{(m-1)^2 + 0^2}},\rho_{t}^{\sqrt{(m-1)^2 + 1^2}},\cdots,\rho_{t}^{\sqrt{(m-1)^2 + N_{t,h}^2}}\right]\in\mathbb{R}^{N_{t,h}\times N_{t,h}}\\
{\bm T}'_m &= {\rm Tz}\left[\rho_{r}^{\sqrt{(m-1)^2 + 0^2}},\rho_{r}^{\sqrt{(m-1)^2 + 1^2}},\cdots,\rho_{r}^{\sqrt{(m-1)^2 + N_{r,h}^2}}\right]\in\mathbb{R}^{N_{r,h}\times N_{r,h}}.
\end{split}
\eeq
\setcounter{equation}{\value{mytempeqncnt}}
\end{figure*}

Consider a rectangular shape of transmit antenna arrays. Suppose that the minimum distance of adjacent antennas is $\delta$. The antenna index is allocated from top-left antenna to bottom-right antenna, i.e., Zig-Zag. Similarly, we index the receive antennas. Following the antenna indices, we first construct a uncorrelated, spatial-domain MIMO channel matrix ${\bm H}_{\rm iid}(n)\in\mathbb{C}^{N_r\times N_t}$. The correlation is then simply characterized by correlation factors $\rho_t$ and $\rho_r$. The factors $\rho_t$ and $\rho_r$ represent the correlation strength between the adjacent antennas separated by $\delta$ at the transmitter and receiver, respectively. Using $\rho_t$ and $\rho_r$ and the fact that spatial correlation is inversely proportional to the distance $\delta$ between antennas \cite{KeScPeMoFr02}, the ${\bm R}_r$ and ${\bm R}_t$ of the 2-D antennas can be modeled as follows: %are modeled from the geometric structure using the distance between antennas and the corresponding correlation, we can derive the correlation model as follows:
\beqn
{\bm R}_{\rm t} &=& {\rm BlkTz}[{\bm T}_1,\cdots,{\bm T}_{N_{t,v}}]\in\mathbb{R}^{N_t\times N_t}\nonumber\\
{\bm R}_{\rm r} &=& {\rm BlkTz}[{\bm T}'_1,\cdots,{\bm T}'_{N_{r,v}}]\in\mathbb{R}^{N_r\times N_r},\nonumber
\eeqn
where ${\bm T}_m$ and ${\bm T}_m'$ are defined at (\ref{Tm}) at the bottom of next page \addtocounter{equation}{1}; and ${\rm BlkTz}[{\bm A}_1,\cdots,{\bm A}_N]$ and ${\rm Tz}[a_1,\cdots,a_N]$ produce a symmetric block Toeplitz and Toeplitz matrices as
\beq\nonumber
\bmats
{\bm A}_1& \cdots& {\bm A}_N \\
\vdots & \ddots & \vdots \\
{\bm A}_N& \cdots & {\bm A}_1
\emats
{~\text{and}~}
\bmats
a_1& \cdots& a_N \\
\vdots & \ddots & \vdots \\
a_N& \cdots & a_1
\emats,
\eeq
respectively.

\setcounter{equation}{0}
\section{Proof of Property \ref{prop3}}\label{Appendix:C}
Suppose that the covariance matrix of the original channels is decomposed as $\E\left({\bm h}{\bm h}^H\right)={\bm U}{\bm D}{\bm U}^H$. Let the new channel structure with an arbitrary permutation matrix ${\bm P}$ be ${\bm h}'={\bm P}{\bm h}$. Then the new covariance matrix of the new CSI vector is derived as
\beqn
\E\left({\bm h}'({\bm h}')^{H}\right)&=&\E\left({\bm P}{\bm h}{\bm h}^H{\bm P}^H\right)
={\bm P}\E\left({\bm h}{\bm h}^H\right){\bm P}^H\nonumber\\
&=&{\bm P}{\bm U}{\bm D}{\bm U}^H{\bm P}^H.\label{NewKLT}
\eeqn
From (\ref{NewKLT}), we get the new KLT matrix as $\overline{\bf \Psi}' = \left({\bm P}{\bm U}\right)^H$. Now, using the new KLT matrix, we get the new sparse channel vector ${\bm s}'=\overline{\bf \Psi}'{\bm h}'$. From the property of the permutation matrix that ${\bm P}^H={\bm P}^{-1}$, we get ${\bm s}' = ({\bm P}{\bm U})^H{\bm h}' = ({\bm P}{\bm U})^H{\bm P}{\bm h} = {\bm U}^H{\bm h}$, and can show that the new sparse channels are uncorrelated as follows:
\beq
\E\left({\bm s'}({\bm s}')^H\right) = {\bm U}^H\E\left({\bm h}{\bm h}^H\right){\bm U}= {\bm U}^H {\bm U}{\bm D}{\bm U}^H{\bm U}= {\bm D},\nonumber
\eeq
which implies that the same CSI recovery performance will be achieved regardless of ${\bm P}$.

\setcounter{equation}{0}
\section{Proof of Property \ref{prop1}}\label{Appendix:A}

\begin{IEEEproof}
%
%------long equation
\newcounter{tmpeq1}
\begin{figure*}[b!]
\hrulefill%
\vspace*{4pt}
\normalsize
\setcounter{tmpeq1}{\value{equation}} %
\beqn
\label{nnCorr}%
%\begin{split}
\E\left({\bm h}_n{\bm h}_{n'}^H\right)
\!\!\!\!\!&=&\!\!\!\bmat \E\left(\sum_{i=1}^{N_r} \left({\bm r}_{i}{\bm h}_i^{\rm r}(n)\right){\bm t}_1{\bm t}_1^H \sum_{i=1}^{N_r}\left({\bm h}_i^{{\rm r},H}(n'){\bm r}_i^H\right) \right) & \cdots & \E\left(\sum_{i=1}^{N_r} \left({\bm r}_{i}{\bm h}_i^{\rm r}(n)\right){\bm t}_1{\bm t}_{N_t}^H \sum_{i=1}^{N_r}\left({\bm h}_i^{{\rm r},H}(n'){\bm r}_i^H\right) \right) \\
\vdots & \ddots & \vdots \\
\E\left(\sum_{i=1}^{N_r} \left({\bm r}_{i}{\bm h}_i^{\rm r}(n)\right){\bm t}_{N_t}{\bm t}_1^H \sum_{i=1}^{N_r}\left({\bm h}_i^{{\rm r},H}(n'){\bm r}_i^H\right) \right) & \cdots &
\E\left(\sum_{i=1}^{N_r} \left({\bm r}_{i}{\bm h}_i^{\rm r}(n)\right){\bm t}_{N_t}{\bm t}_{N_t}^H \sum_{i=1}^{N_r}\left({\bm h}_i^{{\rm r},H}(n'){\bm r}_i^H\right) \right)
\emat\nonumber\\
&\overset{(a)}{=}&\!\!\!\bmat \E\left(\sum_{i=1}^{N_r} \left({\bm r}_{i}{\bm h}_i^{\rm r}(n){\bm t}_1{\bm t}_1^H{\bm h}_i^{{\rm r},H}(n'){\bm r}_i^H\right) \right) & \cdots & \E\left(\sum_{i=1}^{N_r} \left({\bm r}_{i}{\bm h}_i^{\rm r}(n){\bm t}_1{\bm t}_{N_t}^H {\bm h}_i^{{\rm r},H}(n'){\bm r}_i^H\right) \right) \\
\vdots & \ddots & \vdots \\
\E\left(\sum_{i=1}^{N_r} \left({\bm r}_{i}{\bm h}_i^{\rm r}(n){\bm t}_{N_t}{\bm t}_1^H {\bm h}_i^{{\rm r},H}(n'){\bm r}_i^H\right) \right) & \cdots &
\E\left(\sum_{i=1}^{N_r} \left({\bm r}_{i}{\bm h}_i^{\rm r}(n){\bm t}_{N_t}{\bm t}_{N_t}^H {\bm h}_i^{{\rm r},H}(n'){\bm r}_i^H\right) \right)
\emat\nonumber\\
&\overset{(b)}{=}&\!\!\!\bmat
c_{nn'}^2 {\bm t}_1^H{\bm t}_1 \sum_{i=1}^{N_r}\left({\bm r}_i{\bm r}_i^H\right) & \cdots & c_{nn'}^2 {\bm t}_{N_t}^H{\bm t}_1 \sum_{i=1}^{N_r}\left({\bm r}_i{\bm r}_i^H\right) \\
\vdots & \ddots & \vdots \\
c_{nn'}^2 {\bm t}_1^H{\bm t}_{N_t} \sum_{i=1}^{N_r}\left({\bm r}_i{\bm r}_i^H\right) & \cdots & c_{nn'}^2 {\bm t}_{N_t}^H{\bm t}_{N_t} \sum_{i=1}^{N_r}\left({\bm r}_i{\bm r}_i^H\right)
\emat%\nonumber\\
=%\!\!\!
c_{nn'}^2
\bmat {\bm t}_1^H{\bm t}_{1} & \cdots & {\bm t}_{N_t}^H{\bm t}_{1}\\
\vdots & \ddots & \vdots \\
{\bm t}_1^H{\bm t}_{N_t} & \cdots & {\bm t}_{N_t}^H{\bm t}_{N_t}
\emat
\otimes \sum_{i=1}^{N_r}\left({\bm r}_i{\bm r}_i^H\right)\nonumber\\
&=&\!\!\!
c_{nn'}^2 \left({\bm R}_{\rm t}\right) \otimes \left({\bm R}_{\rm r}\right)
%\end{split}
\eeqn
\setcounter{equation}{\value{tmpeq1}}
\end{figure*}
%
%------long equation
\newcounter{tmpeq2}
\begin{figure*}[!b]
\hrulefill%
\vspace*{4pt}
\normalsize
\setcounter{tmpeq2}{\value{equation}} %
\setcounter{equation}{1}%
\beq
\label{corr}
%\begin{split}
\E\left({\bm h}{\bm h}^H\right)
\!\!=\!\!
\bmats
\E\left({\bm h}_1{\bm h}_1^H\right) & \cdots & \E\left({\bm h}_1{\bm h}_{N_f}^H\right) \\
\vdots & \ddots & \vdots \\
\E\left({\bm h}_{N_f}{\bm h}_1^H\right) & \cdots & \E\left({\bm h}_{N_f}{\bm h}_{N_f}^H\right)
\emats%\nonumber\\&=&
\!\!=\!\!
\bmats
c_{11}^2 \left({\bm R}_{\rm t}\right) \otimes \left({\bm R}_{\rm r}\right) & \cdots & c_{1N_f}^2 \left({\bm R}_{\rm t}\right) \otimes \left({\bm R}_{\rm r}\right)\\
\vdots & \ddots & \vdots \\
c_{N_f 1}^2 \left({\bm R}_{\rm t}\right) \otimes \left({\bm R}_{\rm r}\right) & \cdots & c_{N_fN_f}^2 \left({\bm R}_{\rm t}\right) \otimes \left({\bm R}_{\rm r}\right)
\emats%\nonumber\\&=&
\!\!=\!\!
\bmats
c_{11}^2&\cdots&c_{1N_f}^2\\
\vdots & \ddots & \vdots \\
c_{N_f1}^2&\cdots&c_{N_fN_f}^2
\emats
\otimes\left({\bm R}_{\rm t} \otimes {\bm R}_{\rm r}\right).
%\end{split}\label{corr}
\eeq
\setcounter{equation}{\value{tmpeq2}}
\end{figure*}
Let express the spatial correlation matrices and the uncorrelated channel matrix as follows: ${\bm R}_{\rm r}^{\frac{1}{2}} = \bmat {\bm r}_1&\cdots&{\bm r}_{N_r} \emat$; ${\bm R}_{\rm t}^{\frac{1}{2}} = \bmat {\bm t}_1&\cdots&{\bm t}_{N_t} \emat $; and ${\bm H}_{\rm iid}(n) = \bmats ({\bm h}_1^{\rm r}(n))^T \cdots ({\bm h}_{N_r}^{\rm r}(n))^T \emats^T$, where ${\bm r}_i\in\mathbb{R}^{N_r\times 1}$ and ${\bm t}_i\in\mathbb{R}^{N_t\times 1}$ are the $i$th column vectors of ${\bm R}_{\rm r}^{\text{\textonehalf}}$ and ${\bm R}_{\rm t}^{\text{\textonehalf}}$, respectively, and ${\bm h}_j^{\rm r}(n)\in\mathbb{C}^{N_t\times 1}$ is the $j$th row vector of ${\bm H}_{\rm iid}(n)$. Then, we can express the channel vector of the $n$th subcarrier as
\beq\nonumber
{\bm h}_n
=
\Big[
\Big(\sum_{i=1}^{N_r} \left({\bm r}_i {\bm h}_i^{\rm r}(n) \right) {\bm t}_1\Big)^T
\cdots
\Big(\sum_{i=1}^{N_r} \left({\bm r}_i {\bm h}_i^{\rm r}(n) \right) {\bm t}_{N_t}\Big)^T
\Big]^T,
\eeq
%${\bm h}_n^T=[(\sum_{i=1}^{N_r} ({\bm r}_i {\bm h}_i^{\rm r}(n) ) {\bm t}_1)^T ~\cdots~(\sum_{i=1}^{N_r} ({\bm r}_i {\bm h}_i^{\rm r}(n)) {\bm t}_{N_t})^T]$,
and derive the cross correlation matrix between the subcarrier $n$ and $n'$ as (\ref{nnCorr}) at the bottom of this page.\addtocounter{equation}{1} Herein, $c_{nn'}^2=\E\left(\sum_{i=1}^{N_r}{\bm h}_i^{\rm r}(n){\bm h}_i^{{\rm r},H}(n')\right)/\left(N_rN_t\right)$ is the correlation of the channels of frequency $n$ and $n'$. In (\ref{nnCorr}), (a) follows since ${\bm h}_i^{\rm r}(n)$ and ${\bm h}_j^{\rm r}(n')$ are uncorrelated if $i\neq j$, $\forall n, n'\in\{1,\ldots,N_f\}$, (b) follows from the independence of all elements of ${\bm h}_i^{\rm r}(n)$, so that $\E\left({\bm h}_i^{\rm r}(n){\bm t}_a {\bm t}_b^H {\bm h}_i^{{\rm r},H}(n')\right) =c_{nn'}^2 \tr \left({\bm t}_a{\bm t}_b^H\right)= c_{nn'}^2 {\bm t}_b^H{\bm t}_a$. Using (\ref{nnCorr}), the covariance matrix of the channel is simply rewritten as (\ref{corr}) at the bottom of this page.
\addtocounter{equation}{1}

Now, based on the definition of the frequency domain channels, we derive $c_{nn'}^2$ as follows:
\beq
\begin{split}
c_{nn'}^2&=\frac{1}{N_rN_t}\E\left(\sum_{i=1}^{N_r}{\bm h}_i^{\rm r}(n){\bm h}_i^{{\rm r},H}(n')\right)\nonumber\\
&= \E \left(\left[{\bm F}_L \left({\bm  h}_{r,t}\odot\sqrt{\bm d}\right)\right]_{n}\left[{\bm F}_L \left({\bm  h}_{r,t}\odot\sqrt{\bm d}\right)\right]_{n'}^*\right)\nonumber\\
&= \E \left( {\bm f}_n^{\rm r} \left({\bm  h}_{r,t}\odot\sqrt{\bm d}\right)\left(\sqrt{{\bm d}^H}\odot{\bm  h}_{r,t}^H\right){\bm f}_{n'}^{{\rm r},H}\right)\nonumber\\
&= \tr \left( {\bm f}_{n'}^{{\rm r},H}{\bm f}_{n}^{\rm r} \E \left({\bm h}_{r,t} \odot \sqrt{\bm d}\sqrt{{\bm d}^H}\odot {\bm h}_{r,t}^H \right)   \right)\nonumber\\
&= \tr \left( {\bm f}_{n'}^{{\rm r},H}{\bm f}_{n}^{\rm r} \diag({\bm d})\;\sigma_h^2\right),\nonumber%\\
%&=& \tr \left( {\bm f}_{1}^{{\rm r},H}{\bm f}_{n-n'+1}^{\rm r} \diag({\bm d})\right)
\end{split}
\eeq
where ${\bm h}_{r,t}\in\mathbb{C}^{L\times1}$ is $L$-by-$1$ complex normal distributed random variable, i.e., ${\bm h}_{r,t}\sim\mathcal{CN}(0,1)$, for realizing the time domain channels from transmit antenna $t$ to the receive antenna $r$; $\odot$ represents the elementwise product; and $[{\bm a}]_n$ and $[{\bm a}]_n^*$ are the $n$th element of a vector ${\bm a}$ and its complex conjugate, respectively. Since ${\bm f}_{n'}^{{\rm r},H}{\bm f}_{n}^{{\rm r}} = {\bm f}_{n_1}^{{\rm r},H}{\bm f}_{n_2}^{{\rm r}}$ for any $n_1$ and $n_2$ such that $|n_2-n_1|=|n'-n|$, by denoting $c_{n}^2=c_{1n}^2$, we can simply rewrite the first term in (\ref{corr}) as follows:
\beq\nonumber
\bmats
c_{11}^2&\cdots&c_{1N_f}^2\\
\vdots & \ddots & \vdots \\
c_{N_f1}^2&\cdots&c_{N_fN_f}^2
\emats
= {\rm Tz}[c_{1}^2,\cdots,c_{N_f}^2],
\eeq
where $c_{n}^2= \sigma_h^2 \tr \left( {\bm f}_{1}^{{\rm r},H}{\bm f}_{n}^{\rm r} \diag({\bm d})\right)$. This completes the proof.

\end{IEEEproof}

\setcounter{equation}{0}
\section{NMSE Derivation in (\ref{eqn:deltaM})}\label{Appendix:E}
%\begin{IEEEproof}
The selection matrix ${\bm S}$ is given by ${\bm S}_{\rm PCA}(M)$, and the corresponding recovered channel is $\widetilde{\bm h} = {\bm C}_{\bm h}^{\text{\textonehalf}} {\bm S}^\dag {\bm S} {\bm a}$. Note that this is in line with the definition in (\ref{CSr}), where the representation matrix ${\bf \Psi} = {\bm U}^H$ is the singular matrix comprising the eigenvectors of ${\bm C}_{\bm h}$, and the sparse signal ${\bm s} = {\bm D}^{\text{\textonehalf}} {\bm a}$ is scaled according to the diagonal singular values of ${\bm C}_{\bm h}$. Thus, we can derive the NMSE as follows:
\beq\nonumber
\begin{split}
\delta(M) % &= \frac{\E\|{\bm h}-\widetilde{\bm h}\|_2^2}{\E\|{\bm  h}\|_2^2} \\
&={{\rm tr}\left(\E\big( ( {\bm h}-\widetilde{\bm h} ) ( {\bm h}-\widetilde{\bm h})^H \big)\right)}\Big/{{\rm tr}\left(\E\left({\bm h} {\bm  h}^H \right) \right)}  \\
&={ {\rm tr}\left(\! {\bm C}_{\bm h}^{\frac{1}{2}}\!\! \left( {\bm I} \!-\! {\bm S}^\dag {\bm S}  \right) \!\E\!\left({\bm a} {\bm a}^H \right) \!\!\left( {\bm I} \!-\! {\bm S}^\dag {\bm S}  \right)^H \!\!{\bm C}_{\bm h}^{\frac{1}{2}}  \right) }\!\Big/\!{{\rm tr}\left( {\bm C}_{\bm h} \right)} \\
&= {\left({\rm tr}\left({\bm D} \right)- {\rm tr}\left({\bm S} {\bm D} \right)\right)}\big/{{\rm tr}\left({\bm D} \right)}.
\end{split}
\eeq
%\end{IEEEproof}



\begin{thebibliography}{10}
\providecommand{\url}[1]{#1}
\csname url@samestyle\endcsname
\providecommand{\newblock}{\relax}
\providecommand{\bibinfo}[2]{#2}
\providecommand{\BIBentrySTDinterwordspacing}{\spaceskip=0pt\relax}
\providecommand{\BIBentryALTinterwordstretchfactor}{4}
\providecommand{\BIBentryALTinterwordspacing}{\spaceskip=\fontdimen2\font plus
\BIBentryALTinterwordstretchfactor\fontdimen3\font minus
  \fontdimen4\font\relax}
\providecommand{\BIBforeignlanguage}[2]{{%
\expandafter\ifx\csname l@#1\endcsname\relax
\typeout{** WARNING: IEEEtran.bst: No hyphenation pattern has been}%
\typeout{** loaded for the language `#1'. Using the pattern for}%
\typeout{** the default language instead.}%
\else
\language=\csname l@#1\endcsname
\fi
#2}}
\providecommand{\BIBdecl}{\relax}
\BIBdecl

\bibitem{JoSu14GC}
J.~Joung and S.~Sun, ``{SCF}: {Sparse} channel-state-information feedback using
  {Karhunen}-{L}o{\`{e}}ve transform,'' in \emph{Proc. {IEEE} Global Commun.
  Conf. ({GLOBECOM})}, Austin, TX, USA, Dec. 2014, pp. 399--404.

\bibitem{ZhGi02}
S.~Zhou and G.~B. Giannakis, ``Optimal transmitter eigen-beamforming and
  space-time block coding based on channel mean feedback,'' \emph{{IEEE} Trans.
  Signal Process.}, vol.~50, pp. 2599--2613, Oct. 2002.

\bibitem{SpSwHa04}
Q.~H. Spencer, A.~L. Swindlehurst, and M.~Haardt, ``Zero-forcing methods for
  downlink spatial multiplexing in multi-user {MIMO} channels,'' \emph{{IEEE}
  Trans. Signal Process.}, vol.~52, pp. 461--471, Feb. 2004.

\bibitem{ChMu04}
L.-U. Choi and R.~Murch, ``A transmit preprocessing technique for multiuser
  {MIMO} systems using a decomposition approach,'' \emph{{IEEE} Trans. Wireless
  Commun.}, pp. 20--24, Jan. 2004.

\bibitem{JoKiLiJaShChChLe06}
J.~Joung, E.~Y. Kim, S.~H. Lim, Y.-U. Jang, W.-Y. Shin, S.-Y. Chung, J.~Chun,
  and Y.~H. Lee, ``Capacity evaluation of various multiuser {MIMO} schemes in
  downlink cellular environments,'' in \emph{Proc. {IEEE} Int. Symp. on
  Personal, Indoor and Mobile Radio Commun. ({PIMRC})}, Helsinki, Finland, Sep.
  2006.

\bibitem{SaTaSa07}
M.~Sadek, A.~Tarighat, and A.~H. Sayed, ``A leakage-based precoding scheme for
  downlink multi-user {MIMO} channels,'' \emph{{IEEE} Trans. Wireless Commun.},
  pp. 1711--1721, May 2007.

\bibitem{LiChGeSaShZh12}
L.~Liu, R.~Chen, S.~Geirhofer, K.~Sayana, Z.~Shi, and Y.~Zhou, ``Downlink
  {MIMO} in {LTE}-advanced: {SU-MIMO} vs. {MU-MIMO},'' \emph{{IEEE} Commun.
  Mag.}, vol.~50, no.~2, pp. 140--147, Feb. 2012.

\bibitem{11n}
\emph{IEEE Std 802.11n-2009}, NY, USA, IEEE Std.

\bibitem{11ac}
\emph{IEEE Std 802.11ac/D7.0, Sept 2013}, NY, USA, IEEE Std.

\bibitem{GuKa13}
F.~Verbeyst and M.~Bossche, ``Real-time and optimal {PA} characterization
  speeds up {PA} design,'' in \emph{34th European Microwave Conference},
  Amsterdam, Netherlands, Oct. 2004, pp. 431--434.

\bibitem{BjHoKoDe13}
\BIBentryALTinterwordspacing
E.~Bj{\"{o}}rnson, J.~Hoydis, M.~Kountouris, and M.~Debbah, ``Massive {MIMO}
  systems with non-ideal hardware: {Energy} efficiency, estimation, and
  capacity limits.'' [Online]. Available: \url{http://arxiv.org/abs/1304.0553}
\BIBentrySTDinterwordspacing

\bibitem{RuPeLaLaMaEdTu13}
F.~Rusek, D.~Persson, B.~K. Lau, E.~G. Larsson, T.~L. Marzetta, O.~Edfors, and
  F.~Tufvesson, ``Scaling up {MIMO}: {Opportunities} and challenges with very
  large arrays,'' \emph{{IEEE} Trans. Signal Process.}, vol.~30, pp. 40--60,
  Jan. 2013.

\bibitem{JaJoShJe11}
Y.-U. Jang, J.~Joung, W.-Y. Shin, and E.-R. Jeong, ``Frame design and
  throughput evaluation for practical multiuser {MIMO OFDMA} systems,''
  \emph{{IEEE} Trans. Veh. Technol.}, vol.~60, no.~7, pp. 3127--3141, Sep.
  2011.

\bibitem{YiVoThLoGh14}
\BIBentryALTinterwordspacing
D.~Ying, F.~W. Vook, T.~A. Thomas, D.~J. Love, and A.~Ghosh, ``Kronecker
  product correlation model and limited feedback codebook design in a {3D}
  channel model.'' [Online]. Available:
  \url{http://arxiv.org/pdf/1401.2952v1.pdf}
\BIBentrySTDinterwordspacing

\bibitem{GhRaZe10}
S.~Ghosh, B.~D. Rao, and J.~R. Zeidler, ``Outage-efficient strategies for
  multiuser {MIMO} networks with channel distribution information,''
  \emph{{IEEE} Trans. Signal Process.}, vol.~58, pp. 6312--6324, Dec. 2010.

\bibitem{MaEr10}
B.~Makki and T.~Eriksson, ``Efficient channel quality feedback signaling using
  transform coding and bit allocation,'' in \emph{Proc. {IEEE} Veh. Technol.
  Conf. ({VTC})}, Taipei, Taiwan, May 2010, pp. 1--5.

\bibitem{R1-090618}
``Codebook design for 8 {Tx} transmission in {LTE-A},'' Samsung, Athens,
  Greece, Tech. Rep. R1-090618, Feb. 2009.

\bibitem{WaCoDeSl12}
S.~Wagner, R.~Couillet, M.~Debbah, and D.~Slock, ``Large system analysis of
  linear precoding in correlated {MISO} broadcast channels under limited
  feedback,'' \emph{{IEEE} Trans. Inf. Theory}, vol.~58, no.~7, pp. 4509--4537,
  Jul. 2012.

\bibitem{JoChSu14}
J.~Joung, Y.~K. Chia, and S.~Sun, ``Energy-efficient, large-scale
  distributed-antenna system ({L-DAS}) for multiple users,'' \emph{{IEEE} J.
  Sel. Topics Signal Process.}, vol.~8, pp. 954--965, Oct. 2014.

\bibitem{LeSh14}
B.~Lee and B.~Shim, ``An efficient feedback compression for large-scale {MIMO}
  systems,'' in \emph{Proc. {IEEE} Veh. Technol. Conf. ({VTC-Spring})}, Seoul,
  Korea, May 2014.

\bibitem{LeChSeLoSh15}
\BIBentryALTinterwordspacing
B.~Lee, J.~Choi, J.~yun Seol, D.~J. Love, and B.~Shim, ``Antenna grouping based
  feedback compression for fdd-based massive mimo systems.'' [Online].
  Available: \url{http://http://arxiv.org/abs/1408.6009v2}
\BIBentrySTDinterwordspacing

\bibitem{Jo86Book}
I.~T. Jolliffe, \emph{Principal Component Analysis}, 2nd~ed.\hskip 1em plus
  0.5em minus 0.4em\relax New York: Springer-Verlag, 1986.

\bibitem{JoNeRiSe10}
M.~Journ{\'{e}}e, Y.~Nesterov, P.~Richt{\'{a}}rik, and R.~Sepulchre,
  ``Generalized power method for sparse principal component analysis,''
  \emph{J. Mach. Learn. Res.}, vol.~11, pp. 517--553, Feb. 2010.

\bibitem{LiCh93}
K.~Liu and C.-T. Chiu, ``Unified parallel lattice structures for time-recursive
  discrete cosine/sine/hartley transforms,'' \emph{{IEEE} Trans. Signal
  Process.}, vol.~41, pp. 1357--1377, Mar. 1993.

\bibitem{KuKuTi12}
P.-H. Kuo, H.~T. Kung, and P.-A. Ting, ``Compressive sensing based channel
  feedback protocols for spatially-correlated massive antenna arrays,'' in
  \emph{Proc. {IEEE} Wireless Commun. Netw. Conf. ({WCNC})}, Paris, France,
  Apr. 2012, pp. 492--497.

\bibitem{GwKuVl12}
Y.~Gwon, H.~T. Kung, and D.~Vlah, ``Compressive sensing with optimal
  sparsifying basis and applications in spectrum sensing,'' in \emph{Proc.
  {IEEE} Global Commun. Conf. ({GLOBECOM})}, Anaheim, CA, USA, Dec. 2012, pp.
  5386--5391.

\bibitem{KuJoSu15}
E.~Kurniawan, J.~Joung, and S.~Sun, ``Limited feedback scheme for massive
  {MIMO} in mobile multiuser {FDD} systems,'' in \emph{Proc. {IEEE} Int. Conf.
  Commun. ({ICC})}, London, UK, Jun. 2015.

\bibitem{CaRoTa06}
E.~J. Cand{\`{e}}s, J.~Romberg, and T.~Tao, ``Robust uncertainty principles:
  Exact signal reconstruction from highly incomplete frequency information,''
  \emph{{IEEE} Trans. Inf. Theory}, vol.~52, no.~2, pp. 489--509, Feb. 2006.

\bibitem{MaQuMuRoWiZo09}
R.~Masiero, G.~Quer, D.~Munaretto, M.~Rossi, J.~Widmer, and M.~Zorzi, ``Data
  acquisition through joint compressive sensing and principal component
  analysis,'' in \emph{Proc. {IEEE} Global Commun. Conf. ({GLOBECOM})},
  Honolulu, Hawaii, USA, Dec. 2009, pp. 1--6.

\bibitem{HuLi98}
Y.~Hua and W.~Liu, ``Generalized {Karhunen}-{L}{o\`{e}}ve transform,''
  \emph{{IEEE} Signal Process. Lett.}, vol.~5, no.~6, pp. 141--142, Jun. 1998.

\bibitem{KeScPeMoFr02}
J.~P. Kermoal, L.~Schumacher, K.~I. Pedersen, P.~E. Mogensen, and
  F.~Frederiksen, ``A stochastic {MIMO} radio channel model with experimental
  validation,'' \emph{{IEEE} J. Sel. Areas Commun.}, vol.~20, pp. 1211--1226,
  Aug. 2002.

\bibitem{AdNaAhCa13}
A.~Adhikary, J.~Nam, J.~Y. Ahn, and G.~Caire, ``Joint spatial division and
  muliplexing---{The} large-scale array regime,'' \emph{{IEEE} Trans. Inf.
  Theory}, vol.~59, no.~10, pp. 6441--6463, Oct. 2013.

\bibitem{PrSa07book}
J.~G. Proakis and M.~Salehi, \emph{Digital communications}, five~ed.\hskip 1em
  plus 0.5em minus 0.4em\relax New York,: McGraw-Hill, 2007.

\bibitem{MaHaKo13}
T.~Matsumoto, Y.~Hatakawa, and S.~Konishi, ``Experimental performance
  evaluation of time-domain {CSI} compression scheme for multiuser {MIMO},'' in
  \emph{Proc. {IEEE} Asia Pacific Wirelss Communicatins Symposium ({APWCS})},
  Seoul, Korea, Aug. 2013, pp. 327--331.

\bibitem{Bo78Book}
C.~d. Boor, \emph{A Practical Guide to Splines}, revised~ed., ser. Applied
  Mathematical Sciences.\hskip 1em plus 0.5em minus 0.4em\relax New York:
  Springer-Verlag, 2001.

\end{thebibliography}
\end{document}